\documentclass[orivec,runningheads]{llncs}
\usepackage[paperheight=225mm,paperwidth=155mm,textwidth=12.2cm,textheight=19.3cm,centering]{geometry}
\usepackage[hidelinks]{hyperref}
\usepackage{underscore}
\usepackage{color}

\usepackage{mathabx}
\usepackage{graphicx}
\def\orcidID#1{{\href{http://orcid.org/#1}{\includegraphics{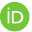}}}}
\DeclareSymbolFont{frenchscript}{OMS}{ztmcm}{m}{n}
\DeclareMathSymbol{\Pow}{\mathord}{frenchscript}{80}   
\newtheorem{defi}{Definition}
\newtheorem{theo}{Theorem}
\newtheorem{prop}{Proposition}
\newtheorem{obse}{Observation}

\newenvironment{definitionR}[1]{\begin{trivlist} \item[]\begin{defi}\label{df:#1}\rm}
                           {\end{defi}\end{trivlist}}
\newenvironment{theoremR}[1]{\begin{trivlist} \item[]\begin{theo}\label{thm:#1}\rm}
                           {\end{theo}\end{trivlist}}
\newenvironment{theoremRc}[2]{\begin{trivlist} \item[]\begin{theo}[#2]\label{thm:#1}\rm}
                           {\end{theo}\end{trivlist}}
\newenvironment{propositionR}[1]{\begin{trivlist} \item[]\begin{prop}\label{pr:#1}\rm}
                           {\end{prop}\end{trivlist}}
\newenvironment{observationR}[1]{\begin{trivlist} \item[]\begin{obse}\label{obs:#1}\rm}
                           {\end{obse}\end{trivlist}}

\newcommand{\df}[1]{Def.~\ref{df:#1}}
\newcommand{\thm}[1]{Thm.~\ref{thm:#1}}
\newcommand{\pr}[1]{Prop.~\ref{pr:#1}}
\newcommand{\sect}[1]{Section~\ref{sec:#1}}
\newcommand{\fig}[1]{Figure~\ref{fig:#1}}
\newcommand{\C}{\mathcal{T}}          
\newcommand{\acts}{{\it acts}}
\newcommand{\states}{{\it states}}
\newcommand{\start}{{\it start}}
\newcommand{\steps}{{\it steps}}
\newcommand{\pa}{{\it part}}
\newcommand{\ina}{{\it in}}
\newcommand{\out}{{\it out}}
\newcommand{\inta}{{\it int}}
\newcommand{\local}{{\it local}}
\newcommand{\ext}{{\it ext}}
\newcommand{\length}{{\it length}}
\newcommand{\sched}{{\it sched}}
\newcommand{\trace}{{\it trace}}
\newcommand{\traces}{{\it fintraces}}
\newcommand{\fairtraces}{{\it fairtraces}}
\newcommand{\qtraces}{{\it qtraces}}
\newcommand{\reward}{{\it reward}}
\begin{document}
\let\doi\undefined

\title{Fair Must Testing for I/O Automata}
\author{Rob van Glabbeek\orcidID{0000-0003-4712-7423}}
\authorrunning{R.J. van Glabbeek}
\institute{School of Comp.\ Sc.\ and Engineering, Univ.\ of New South Wales, Sydney, Australia
\email{rvg@cs.stanford.edu}}

\maketitle

\begin{abstract}
The concept of must testing is naturally parametrised with a chosen
completeness criterion or fairness assumption. When taking weak
fairness as used in I/O automata, I show that it characterises exactly
the fair preorder on I/O automata as defined by Lynch \& Tuttle.

\end{abstract}

\noindent
{\it This paper is dedicated to Frits Vaandrager at the occasion of his 60th birthday.
I fondly remember my days at CWI as a starting computer scientist, sharing an office with Frits.
Here I had the rare privilege of sharing all my ideas with Frits at the time they were formed, and
receiving instantaneous meaningful feedback. This feedback has had a great impact on my work.

I take the opportunity to also pass best wishes and warmest thoughts to Frits from Ursula Goltz,
whom I am visiting while finishing this paper. My joint work with Ursula was inspired by my
work with Frits on connecting Petri nets and process algebra.
}

\section{Introduction}

May- and must-testing was proposed by De Nicola \& Hennessy in \cite{DH84}. It yields semantic
equivalences where two processes or automata are distinguished if and only if they react differently on certain
tests. The tests are processes that additionally feature success states. Such a test $T$ is applied
to a process $A$ by taking the CCS parallel composition $T|A$, and implicitly applying a CCS
restriction operator to it that removes the remnants of uncompleted communication attempts. The
outcome of applying $T$ to $A$ is deemed successful if and only if this composition yields a process
that may, respectively must, reach a success state.
It is trivial to recast this definition of may- and must-testing equivalence using the CSP parallel
composition $\|$ \cite{Ho85} instead of the one from CCS.

I/O automata \cite{LT89} are a model of concurrency that distinguishes output actions, which are
under the control of a given automaton, from input actions, which are stimuli from the environment
on which an automaton might react. The parallel composition $\|$ of I/O automata, exactly like the
one of CSP, imposes synchronisation on actions the composed automata have in common. However, it
allows forming the composition $A\|B$ only when $A$ and $B$ have no output actions in common.
This makes it impossible to synchronise on actions $c$ where both $A$ and $B$ have the option not to
allow $c$ in certain states.

Must testing equivalence for CCS and CSP partially discerns \emph{branching time}, in the sense that
is distinguishes the processes $\tau.(a+b)$ and $\tau.a + \tau.b$ displayed in \fig{testing}.
This is not the case for I/O automata, as the synchronisations between test and tested automaton
that are necessary to make such distinctions are ruled out by the restriction described above.

It is not a priori clear how a given process or automaton \emph{must} reach a success state.
For all we know it might stay in its initial state and never take any transition leading to
this success state. To this end one must employ an assumption saying that under appropriate
circumstances certain enabled transitions will indeed be taken. Such an assumption is called a
\emph{completeness criterion} \cite{vG19}. The theory of testing from \cite{DH84} implicitly employs
a default completeness criterion that in \cite{GH19} is called \emph{progress}.
However, one can parameterise the notion of must testing by the choice of any completeness
criterion, such as the many notions of \emph{fairness} classified in \cite{GH19}.

Lynch \& Tuttle \cite{LT89} defined a trace and a fair preorder on I/O automata, which were meant to
reason about safety and liveness properties, respectively, just like the may- and must testing
preorders of \cite{DH84}. Unsurprisingly, as formally shown in \sect{may} of this paper, the trace
preorder on I/O automata is characterised exactly by may testing. Segala \cite{Se97} has
studied must-testing on I/O automata, employing the default completeness criterion,
and found that on a large class of I/O automata it characterises the \emph{quiescent trace
preorder} of Vaandrager \cite{Va91}. It does not exactly characterise the fair preorder, however.

In my analysis this is due to the choice of progress as the completeness criterion employed for must
testing, whereas the fair preorder of I/O automata is based on a form of weak fairness.
In this work I study must testing on I/O automata based on the same form of weak fairness,
and find that it characterises the fair preorder exactly.

Although I refer to must-testing with fairness as the chosen completeness criterion as \emph{fair must testing},
it should not be confused with the notion of \emph{fair testing} employed in \cite{BRV95,NC95}.
The latter is also known as \emph{should testing}. It incorporates a concept of fairness that is much
stronger than the notion of fairness from I/O automata, called \emph{full fairness} in \cite{GH19}.

In \cite{vG19b} another mode of testing was proposed, called \emph{reward testing}.
Reward-testing equivalence combines the distinguishing power of may as well as must testing,
and additionally makes some useful distinctions between processes that are missed by both may and
must testing \cite{vG19b}. As for must testing, its definition is naturally parametrised by a
completeness criterion. When applied to I/O automata, using as completeness criterion the form of
fairness that is native to I/O automata, it turns out that reward testing is not stronger than must
testing, and also characterises the fair preorder.

\newpage

\section{I/O automata}\label{sec:IO}

An I/O automaton is a labelled transition system equipped with a nonempty set of start states,
with each action that may appear as transition label classified as an input, an output or an
internal action. Input actions are under the control of the environment of the automaton,
whereas output and internal actions, together called \emph{locally-controlled} actions,
are under the control of the automaton itself.
I/O automata are \emph{input enabled}, meaning that in each state each input action of the automaton
can be performed. This indicates that the environment may perform such actions regardless of the state of
the automaton; an input transition merely indicates how the automaton reacts on such an event.
To model that certain input actions have no effect in certain states, one uses self-loops.

I/O automata employ a partition of the locally-controlled actions into \emph{tasks} to indicate which
sequences of transitions denote \emph{fair} runs. A run is fair unless it has a
suffix on which some task is enabled in every state, yet never taken.

\begin{definitionR}{IO}
An \emph{input/output automaton} (or \emph{I/O automaton}) $A$ is a tuple
$(\acts(A),\states(A),\start(A),\steps(A),\pa(A))$ with\vspace{-1ex}
\begin{itemize}
\item $\acts(A)$ a set of \emph{actions}, partitioned into three sets $\ina(A)$,
  $\out(A)$ and $\inta(A)$ of \emph{input actions}, \emph{output actions} and
  \emph{internal actions}, respectively,
\item $\states(A)$ a set of \emph{states},
\item $\start(A) \subseteq \states(A)$ a nonempty set of start states,
\item $\steps(A) \subseteq \states(A) \times \acts(A) \times \states(A)$
  a \emph{transition relation} with the property that
  $\forall s \in \states(A).\;\forall a \in \ina(A).\; \exists (s,a,s')\in\steps(A)$, and
\item $\pa(A) \subseteq \Pow(\local(A))$ a partition of the set
  $\local(A) := \out(A) \cup \inta(A)$ of \emph{locally-controlled actions} of $A$ into \emph{tasks}.
\vspace{-1ex}
\end{itemize}
Let $\ext(A) := \ina(A) \cup \out(A)$ be the set of \emph{external actions} of $A$.
\end{definitionR}
An action $a \in \acts(A)$ is \emph{enabled} in a state $s \in \states(A)$ if $\exists (s,a,s')\in\steps(A)$.
A task $\C \in \pa(A)$ is \emph{enabled} in $s$ if some action $a \in \C$ is enabled is $s$.

\begin{definitionR}{fairness}
An \emph{execution} of an I/O automaton $A$ is an alternating sequence $\alpha = s_0,a_1,s_1,a_2,\dots$
of states and actions, either being infinite or ending with a state, such that $s_0 \in \start(A)$
and $(s_{i},a_{i+1},s_{i+1})\in\steps(A)$ for all $i< \length(\alpha)$.
Here $\length(\alpha) \in \bbbn\cup \{\infty\}$ denotes the number of action occurrences in $\alpha$.
The sequence $a_1,a_2,\dots$ obtained by dropping all states from $\alpha$ is called $\sched(\alpha)$.
An execution $\alpha$ of $A$ is \emph{fair} if, for each suffix
$\alpha' = s_k,a_{k+1},s_{k+1},a_{k+2},\dots$ of $\alpha$
(with $k \in \bbbn \wedge k \leq \length(\alpha)$) and each task $\C \in \pa(A)$,
if $\C$ is enabled in each state of $\alpha'$, then $\alpha'$ contains an action from $\C$.
\end{definitionR}
In \cite{LT89} two semantic preorders are defined on I/O automata, here called $\sqsubseteq_T$ and
$\sqsubseteq_F$, the \emph{trace} and the \emph{fair preorder}.
In \cite{LT89} $S \sqsubseteq_T I$ and $S \sqsubseteq_F I$ are denoted
``\emph{$I$ implements $S$}'' and ``\emph{$I$ solves $S$}'', respectively.
Here $S$ is an I/O automaton that is (a step closer to) the specification of a problem, and $I$ one
that is (a step closer to) its implementation.
The preorder $\sqsubseteq_T$ is meant to reason about safety properties: if $S \sqsubseteq_T I$ then
$I$ has any safety property that $S$ has. In the same way,\pagebreak[3] $\sqsubseteq_F$ is for
reasoning about liveness properties. In \cite{Se97} and much subsequent work $S \sqsubseteq_F I$ is
written as $I \sqsubseteq_F S$. Here I put $I$ on the right, so as to orient the refinement symbol
$\sqsubseteq$ in the way used in CSP \cite{Ho85}, and in the theory of testing~\cite{DH84}.

I/O automata are a typed model of concurrency, in the sense that two automata will be compared only
when they have the same input and output actions.

\begin{definitionR}{preorders}
Let $\trace(\alpha)$ be the finite or infinite sequence of external actions resulting from dropping
all internal actions in $\sched(\alpha)$,
and let $\traces(A)$ be the set $\{\trace(\alpha) \mid \mbox{$\alpha$ is a finite execution of $A$}\}$.
Likewise $\fairtraces(A) := \{\trace(\alpha) \mid \mbox{$\alpha$ is a fair execution of $A$}\}$.
Now
\[\begin{array}{c@{~~:\Leftrightarrow~~\ina(S)\mathbin=\ina(I) \wedge \out(S)\mathbin=\out(I) \wedge}c}
S \sqsubseteq_T I & \traces(I) \subseteq \traces(S)
\\[1ex]
S \sqsubseteq_F I & \fairtraces(I) \subseteq \fairtraces(S)\makebox[0pt][l]{\,.}
\end{array}\]
One writes $A \equiv_T B$ if $A \sqsubseteq_T B \wedge B \sqsubseteq_T A$, and similarly for $\equiv_F$.
\end{definitionR}
By \cite[Thm.~6.1]{GH19} each finite execution can be extended into a fair execution.
As a consequence, $A \sqsubseteq_F B \Rightarrow A \sqsubseteq_T B$.

The parallel composition of I/O automata \cite{LT89} is similar to the one of CSP \cite{Ho85}:
participating automata $A_i$ and $A_j$ synchronise on actions in $\acts(A_i) \cap \acts(A_j)$,
while for the rest allowing arbitrary interleaving. However, it is defined only when the participating
automata have no output actions in common.

\begin{definitionR}{composition}
A collection $\{A_i\}_{i\in I}$ of I/O automata is \emph{strongly compatible} if\vspace{-1ex}
\begin{itemize}
\item $\inta(A_i) \cap \acts(A_j) = \emptyset$ for all $i,j\in I$ with $i \neq j$, and
\item $\out(A_i) \cap \out(A_j) = \emptyset$ for all $i,j\in I$ with $i \neq j$,
\item no action is contained in infinitely many sets $\acts(A_i)$.
\end{itemize}
The \emph{composition} $A=\prod_{i \in I}A_i$ of a countable collection $\{A_i\}_{i\in I}$
of strongly compatible I/O automata is defined by\vspace{-1ex}
\begin{itemize}
\item $\inta(A) := \bigcup_{i \in  I} \inta(A_i)$,
\item $\out(A) := \bigcup_{i \in  I} \out(A_i)$,
\item $\ina(A) := \bigcup_{i \in  I} \ina(A_i) - \out(A)$,
\item $\states(A) := \prod_{i \in I} \states(A_i)$,
\item $\start(A) := \prod_{i \in I} \start(A_i)$,
\item $\steps(A)$ is the set of triples $(\vec{s_1} , a , \vec{s_2})$ such that, for all $i \in I$,
if $a \in \acts(A_i)$ then $(\vec{s_1}[i] , a , \vec{s_2}[i]) \in \steps (A_i)$, and
if $a \notin \acts(A_i)$ then $\vec{s_1}[i] = \vec{s_2}[i]$, and
\item $\pa(A) := \bigcup_{i \in  I} \pa(A_i)$.
\end{itemize}
\end{definitionR}
Clearly, composition of I/O automata is associative: when writing $A_1 \| A_2$ for 
$\prod_{i \in \{1,2\}}A_i$ then $(A \| B) \| C \cong A \| (B \| C)$, for some notion of isomorphism
$\cong$, included in $\equiv_T$ and $\equiv_F$.
Moreover, as shown in \cite{LT89},
composition is monotone for $\sqsubseteq_T$ and $\sqsubseteq_F$, or in other words,
$\sqsubseteq_T$ and $\sqsubseteq_F$ are precongruences for composition:
\begin{center}
if $A_i \sqsubseteq_T B_i$ for all $i \in I$, then
$\prod_{i \in I}A_i \sqsubseteq_T \prod_{i \in I}B_i$\makebox[0pt][l]{\,, and}\\[1ex]
if $A_i \sqsubseteq_F B_i$ for all $i \in I$, then
$\prod_{i \in I}A_i \sqsubseteq_F \prod_{i \in I}B_i$\makebox[0pt][l]{\,.}
\end{center}
The first condition of strong compatibility is not a limitation of generality. Each I/O automaton is $\equiv_T$ and
$\equiv_F$-equivalent to the result of bijectively renaming its internal actions. Hence, prior to
composing a collection of automata, one could rename their internal actions to ensure that this
condition is met.  Up to $\equiv_T$ and $\equiv_F$ the composition would be independent on the
choice of these renamings.

\section{Testing preorders}\label{sec:testing}

Testing preorders \cite{DH84} are defined between \emph{automata} $A$, defined as in \df{IO}, but
without the partition $\pa(A)$ and without the distinction between input and output actions, and
therefore also without the input enabling requirement from Item 4.  The parallel composition of
automata is as in \df{composition}, but without the requirement that the participating automata have
no output actions in common.

\begin{definitionR}{automata}
An \emph{automaton} $A$ is a tuple
$(\hspace{-.37pt}\acts(A),\states(A),\start(A),\steps(A)\hspace{-.37pt})$ with\vspace{-2ex}
\begin{itemize}
\item $\acts(A)$ a set of \emph{actions}, partitioned into two sets $\ext(A)$
  and $\inta(A)$ of \emph{external actions} and \emph{internal actions}, respectively,
\item $\states(A)$ a set of \emph{states},
\item $\start(A) \subseteq \states(A)$ a nonempty set of start states, and
\item $\steps(A) \subseteq \states(A) \times \acts(A) \times \states(A)$
  a \emph{transition relation}.
\vspace{-1ex}
\end{itemize}
A collection $\{A_i\}_{i\in I}$ of I/O automata is \emph{compatible} if\vspace{-1ex}
\begin{itemize}
\item $\inta(A_i) \cap \acts(A_j) = \emptyset$ for all $i,j\in I$ with $i \neq j$, and
\item no action is contained in infinitely many sets $\acts(A_i)$.
\end{itemize}
The \emph{composition} $A=\prod_{i \in I}A_i$ of a countable collection $\{A_i\}_{i\in I}$
of compatible I/O automata is defined by\vspace{-1ex}
\begin{itemize}
\item $\inta(A) := \bigcup_{i \in  I} \inta(A_i)$,
\item $\ext(A) := \bigcup_{i \in  I} \ext(A_i)$,
\item $\states(A) := \prod_{i \in I} \states(A_i)$,
\item $\start(A) := \prod_{i \in I} \start(A_i)$, and
\item $\steps(A)$ is the set of triples $(\vec{s_1} , a , \vec{s_2})$ such that, for all $i \in I$,
if $a \in \acts(A_i)$ then $(\vec{s_1}[i] , a , \vec{s_2}[i]) \in \steps (A_i)$, and
if $a \notin \acts(A_i)$ then $\vec{s_1}[i] = \vec{s_2}[i]$.
\end{itemize}
\end{definitionR}
A \emph{test} is such an automaton, but featuring a special external action $w$, not used elsewhere.
This action is used to mark \emph{success states}: those in which $w$ is enabled.
The parallel composition $T\|A$ of a test $T$ and an automaton $A$, if it exists, is itself a test,
and $[T\|A]$ denotes the result of reclassifying all its non-$w$ actions as internal.
An execution of $[T\|A]$ is \emph{successful} iff it contains a success state.
\begin{definitionR}{testing}
An automaton $A$ \emph{may pass} a test $T$, notation $A~\textbf{may}~T$, if $[T\|A]$ has a
successful execution.
It \emph{must pass} $T$, notation $A~\textbf{must}~T$, if each complete execution\footnote{The
original work on must testing \cite{DH84} defined an execution to be complete if it
either is infinite, of ends in a state without outgoing transitions.
Here I will consider the concept of a complete execution as a parameter in the definition of must testing.}
of $[T\|A]$ is successful.
It \emph{should pass} $T$, notation $A~\textbf{should}~T$, if each finite execution of $[T\|A]$ can
be extended into a successful execution.

Write $A \sqsubseteq_{\rm may} B$ if $\ext(A)=\ext(B)$ and
$A~\textbf{may}~T$ implies $B~\textbf{may}~T$ for each test $T$
that is compatible with $A$ and $B$. The preorders $\sqsubseteq_{\rm must}$ and
$\sqsubseteq_{\rm should}$ are defined similarly.
\end{definitionR}
The may- and must-testing preorders stem from \cite{DH84}, whereas should-testing was added
independently in \cite{BRV95} and~\cite{NC95}. I have added the condition $\ext(A)=\ext(B)$ to
obtain preorders that respect the types of automata.
A fourth mode of testing, called \emph{reward testing},
was contributed in \cite{vG19b}. It has no notion of success state, and no action $w$;
instead, each transition of a test $T$ is tagged with a real number, the reward of taking that transition.
A negative reward can be seen as a penalty.
Each transition $(s,a,s')$ of $[T\|A]$ with $a \in \acts(T)$ inherits its reward from the unique
transition of $T$ it projects to; in case  $a \not\in \acts(T)$ it has reward $0$.
The reward $\reward(\alpha)$ of an execution $\alpha$ is the sum of the rewards of the actions in
$\alpha$.\footnote{If $\alpha$ is infinite, its reward can be $+\infty$ or $-\infty$; see
  \cite{vG19b} for a precise definition.}
Now $A \sqsubseteq_{\rm reward} B$ if $\ext(A)=\ext(B)$ and for each test $T$
that is compatible with $A$ and $B$ and for each complete execution $\beta$ of $[T\|B]$
there exists a complete execution $\alpha$ of $[T\|A]$ such that $\reward(\alpha) \leq \reward(\beta)$.

In the original work on testing \cite{DH84,vG19b} the CCS parallel composition $T|A$ was used
instead of the CSP parallel composition $T\|A$; moreover, only those executions consisting solely of
internal actions mattered for the definitions of passing a test. The present approach is equivalent,
in the sense that it trivially gives rise to the same testing preorders.

The may-testing preorder can be regarded as pointing in the opposite direction as the others.
Using CCS notation, one has $\tau.P \sqsubsetneq_{\rm may} \tau.P + \tau.Q$,
yet $\tau.P + \tau.Q \sqsubsetneq_{\rm must} \tau.P$,
$\tau.P + \tau.Q \sqsubsetneq_{\rm should} \tau.P$ and
$\tau.P + \tau.Q \sqsubsetneq_{\rm reward} \tau.P$.
The inverse of the may-testing preorder can be characterised as \emph{survival testing}.
Here a state in which $w$ is enabled is seen as a \emph{failure state} rather than a success state,
and automaton $A$ \emph{survives} test $T$, notation $A~\textbf{surv}~T$, if no execution of $[T\|A]$
passes through a failure state.
Write $A \sqsubseteq_{\rm surv} B$ if $\ext(A)\mathbin=\ext(B)$ and
$A~\textbf{surv}~T$ implies $B~\textbf{surv}~T$ for each test $T$
that is compatible with $A$ and $B$.
By definition, $A \sqsubseteq_{\rm surv} B$ iff $B \sqsubseteq_{\rm may} A$.

The only implications between reward, must and may/survival testing are
\[ A \sqsubseteq_{\rm reward} B  ~~\Rightarrow~~  A \sqsubseteq_{\rm must} B
\qquad\mbox{and}\qquad
   A \sqsubseteq_{\rm reward} B  ~~\Rightarrow~~  A \sqsubseteq_{\rm surv} B\;.\]
Namely, any must test $T$ witnessing $A \not\sqsubseteq_{\rm must} B$ can be coded as a reward test by
assigning a reward $+1$ to all transitions of $T$ leading to a success state (and $0$ to all other transitions).
Likewise any survival test $T$ witnessing $A \not\sqsubseteq_{\rm surv} B$ can be coded as a reward test by
assigning a reward $-1$ to all transitions of $T$ leading to a failure state.

The notions of may- and should-testing are unambiguously defined above, whereas the notions of must-
and reward testing depend on the definition of a complete execution. In \cite{vG19} I posed that
transition systems or automata constitute a good model of distributed systems only in combination
with a \emph{completeness criterion}: a selection of a subset of all executions as complete
executions, modelling complete runs of the represented system.

The default completeness criterion, employed in \cite{DH84,vG19b} for the definition of must- and
reward testing, deems an execution complete if it either is infinite, of ends in deadlock, a state
without outgoing transitions.  Other completeness criteria either classify certain finite
executions that do not end in deadlock as complete, or certain infinite executions as incomplete.

The first possibility was explored in \cite{vG19,GH19} by considering a set $B$ of actions that
might be blocked by the environment in which an automaton is running. Now a finite execution can be
deemed complete if all transitions enabled in its last state have labels from $B$. The system might
stop at such a state if indeed the environment blocks all those actions.
Since in the application to must- and reward testing, all non-$w$ transitions in $[T\|A]$ are labelled with
internal actions, which cannot be blocked by the environment, the above possibility of increasing
the set of finite complete executions does not apply.

The second possibility was extensively explored in \cite{GH19}, where a multitude of completeness
criteria was defined. Most of those can be used as a parameter in the definition of must- and reward testing.
So far, the resulting testing preorders have not been explored.\footnote{The paper \cite{vG22b}
explores these testing preorders; it was written after the present paper.}

\section{Testing preorders for I/O automata}

\begin{figure}
\input{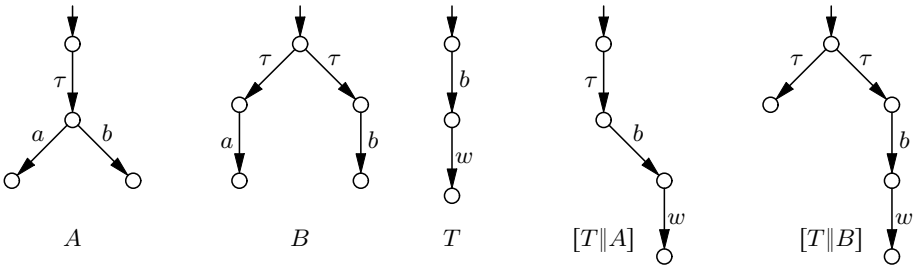}
\centerline{\box\graph}
\caption{Classic example of how branching time is discerned by must testing}
\label{fig:testing}
\end{figure}

Since I/O automata can be seen as special cases of the automata from \sect{testing},
the definitions of \sect{testing} also apply to I/O automata. The condition $\ext(A)=\ext(B)$
should then  be read as $\ina(A)=\ina(B) \wedge \out(A)=\out(B)$.
The only place where it makes an essential difference whether one works with I/O automata or general
automata is in judging compatibility between automata and tests.
Given two I/O automata $A$ and $B$, let $A \sqsubseteq_{\rm must}^{\rm LTS} B$
be defined by first seeing $A$ and $B$ as general automata (by dropping the partitions $\pa(A)$
and $\pa(B)$), and then applying the definitions of \sect{testing}, using the default completeness
criterion. In contrast, let $A \sqsubseteq_{\rm must}^{\it Pr} B$ be defined as \sect{testing}, but
only allowing tests that are themselves I/O automata (seeing the special action $w$ as an output
action), and that are strongly compatible with $A$ and $B$. The superscript {\it Pr} stands for
``progress'', the name given in \cite{GH19} to the default completeness criterion.
The difference between $\sqsubseteq_{\rm must}^{\rm LTS}$ and $\sqsubseteq_{\rm must}^{\it Pr}$ is
illustrated in \fig{testing}.

Here $A$ and $B$ are automata with $\acts(A)=\acts(B)=\{\tau,a,b\}$, and $T$ is a test with
$\acts(T)=\{a,b,w\}$. The short arrows point to start states.
Test $T$ witnesses that $A \not\sqsubseteq_{\rm must}^{\rm LTS} B$, for
$A$ \textbf{must} $T$, yet $\neg (B~{\bf must}~T)$.
Here it is crucial that $a \in \acts(T)$, even though this action labels no
transition of $T$, for otherwise the $a$-transition of $A$ would return in $[T\|A]$ and
one would not obtain $A$ \textbf{must} $T\!$.\linebreak[4]
To see $A$, $B$ and $T$ as I/O automata, one needs to take $\ina(A)=\ina(B)=\ina(T)=\emptyset$,
and thus $a,b\in \out(A)\cap\out(B)\cap\out(T)$. However, this violates the strong compatibility of
$T$ with $A$ and $B$, so that $T$ is disqualified as an appropriate test.
There is no variant of $T$ that is strongly compatible with $A$ and $B$ and yields the same result;
in fact $A \equiv_{\rm must}^{\it Pr} B$.

\section{May testing}\label{sec:may}

For may-testing on I/O automata there is no difference between $\sqsubseteq^{\rm LTS}_{\rm may}$---%
allowing any test that is compatible with $A$ and $B$---and $\sqsubseteq_{\rm may}$---allowing only
tests that are strongly compatible with $A$ and $B$.
These preorders both coincide with the trace preorder $\sqsupseteq_T$.

\begin{theoremR}{may}
$A \sqsubseteq^{\rm LTS}_{\rm may} B$ iff $A \sqsubseteq_{\rm may} B$ iff $B \sqsubseteq_T A$.
\end{theoremR}

\begin{proof}
Suppose $B \sqsubseteq_T A$, i.e., $\ina(A)\mathbin=\ina(B) \wedge \out(A)\mathbin=\out(B)$
and $\traces(A) \subseteq \traces(B)$, and let $T$ be any test compatible with $A$ and $B$.
The automaton $T$ need not be an I/O automaton, and even if it is, it need not be strongly
compatible with $A$ and $B$. It is well-known that $\sqsubseteq_T$ is a
precongruence for composition~\cite{Ho85}, so $\traces(T\|A) \subseteq \traces(T\|B)$.
Since $C$ \textbf{may} $T$ (for any $C$) iff $w$ occurs in a trace $\sigma \in \traces(T\|C)$,
it follows that $A~\textbf{may}~T$ implies $B~\textbf{may}~T$.
Thus $A \sqsubseteq^{\rm LTS}_{\rm may} B$.

That $A \sqsubseteq^{\rm LTS}_{\rm may} B$ implies $A \sqsubseteq_{\rm may} B$ is trivial.

Now suppose $A \sqsubseteq_{\rm may} B$. Then $\ina(A)\mathbin=\ina(B) \wedge \out(A)\mathbin=\out(B)$.
Let $\sigma=a_1 a_2 \dots a_n \in \traces(A)$. Let $T$ be the test automaton\\
\expandafter\ifx\csname graph\endcsname\relax
   \csname newbox\expandafter\endcsname\csname graph\endcsname
\fi
\ifx\graphtemp\undefined
  \csname newdimen\endcsname\graphtemp
\fi
\expandafter\setbox\csname graph\endcsname
 =\vtop{\vskip 0pt\hbox{%
\pdfliteral{
q [] 0 d 1 J 1 j
0.576 w
0.576 w
316.8 -76.824 m
316.8 -79.488222 314.640222 -81.648 311.976 -81.648 c
309.311778 -81.648 307.152 -79.488222 307.152 -76.824 c
307.152 -74.159778 309.311778 -72 311.976 -72 c
314.640222 -72 316.8 -74.159778 316.8 -76.824 c
S
Q
}%
    \graphtemp=.5ex
    \advance\graphtemp by 1.067in
    \rlap{\kern 4.333in\lower\graphtemp\hbox to 0pt{\hss \scriptsize S\hss}}%
\pdfliteral{
q [] 0 d 1 J 1 j
0.576 w
0.072 w
q 0 g
322.776 -72.72 m
315.36 -73.44 l
321.624 -69.336 l
322.776 -72.72 l
B Q
0.576 w
q [3.6] 0 d
315.36 -80.208 m
320.868 -82.116 l
324.61344 -83.41344 328.68 -84.024 333.576 -84.024 c
338.472 -84.024 340.776 -81.72 340.776 -76.824 c
340.776 -71.928 338.472 -69.624 333.576 -69.624 c
328.68 -69.624 324.72864 -70.18848 321.228 -71.388 c
316.08 -73.152 l
S
Q
28.8 -4.824 m
28.8 -7.488222 26.640222 -9.648 23.976 -9.648 c
21.311778 -9.648 19.152 -7.488222 19.152 -4.824 c
19.152 -2.159778 21.311778 0 23.976 0 c
26.640222 0 28.8 -2.159778 28.8 -4.824 c
S
Q
}%
    \graphtemp=.5ex
    \advance\graphtemp by 0.067in
    \rlap{\kern 0.333in\lower\graphtemp\hbox to 0pt{\hss \scriptsize 1\hss}}%
\pdfliteral{
q [] 0 d 1 J 1 j
0.576 w
0.072 w
q 0 g
12.024 -3.024 m
19.224 -4.824 l
12.024 -6.624 l
12.024 -3.024 l
B Q
0.576 w
0 -4.824 m
12.024 -4.824 l
S
76.824 -4.824 m
76.824 -7.488222 74.664222 -9.648 72 -9.648 c
69.335778 -9.648 67.176 -7.488222 67.176 -4.824 c
67.176 -2.159778 69.335778 0 72 0 c
74.664222 0 76.824 -2.159778 76.824 -4.824 c
S
Q
}%
    \graphtemp=.5ex
    \advance\graphtemp by 0.067in
    \rlap{\kern 1.000in\lower\graphtemp\hbox to 0pt{\hss \scriptsize 2\hss}}%
\pdfliteral{
q [] 0 d 1 J 1 j
0.576 w
0.072 w
q 0 g
59.976 -3.024 m
67.176 -4.824 l
59.976 -6.624 l
59.976 -3.024 l
B Q
0.576 w
28.8 -4.824 m
59.976 -4.824 l
S
Q
}%
    \graphtemp=\baselineskip
    \multiply\graphtemp by -1
    \divide\graphtemp by 2
    \advance\graphtemp by .5ex
    \advance\graphtemp by 0.067in
    \rlap{\kern 0.667in\lower\graphtemp\hbox to 0pt{\hss $a_1$\hss}}%
\pdfliteral{
q [] 0 d 1 J 1 j
0.576 w
0.072 w
q 0 g
300.816 -72.144 m
307.368 -75.6 l
299.952 -75.6 l
300.816 -72.144 l
B Q
0.576 w
q [3.6 3.675844] 0 d
28.656 -5.976 m
300.384 -73.872 l
S Q
124.848 -4.824 m
124.848 -7.488222 122.688222 -9.648 120.024 -9.648 c
117.359778 -9.648 115.2 -7.488222 115.2 -4.824 c
115.2 -2.159778 117.359778 0 120.024 0 c
122.688222 0 124.848 -2.159778 124.848 -4.824 c
S
Q
}%
    \graphtemp=.5ex
    \advance\graphtemp by 0.067in
    \rlap{\kern 1.667in\lower\graphtemp\hbox to 0pt{\hss \scriptsize 3\hss}}%
\pdfliteral{
q [] 0 d 1 J 1 j
0.576 w
0.072 w
q 0 g
108 -3.024 m
115.2 -4.824 l
108 -6.624 l
108 -3.024 l
B Q
0.576 w
76.824 -4.824 m
108 -4.824 l
S
Q
}%
    \graphtemp=\baselineskip
    \multiply\graphtemp by -1
    \divide\graphtemp by 2
    \advance\graphtemp by .5ex
    \advance\graphtemp by 0.067in
    \rlap{\kern 1.333in\lower\graphtemp\hbox to 0pt{\hss $a_2$\hss}}%
\pdfliteral{
q [] 0 d 1 J 1 j
0.576 w
0.072 w
q 0 g
301.032 -71.64 m
307.368 -75.456 l
300.024 -75.096 l
301.032 -71.64 l
B Q
0.576 w
q [3.6 3.593104] 0 d
76.608 -6.192 m
300.528 -73.368 l
S Q
220.824 -4.824 m
220.824 -7.488222 218.664222 -9.648 216 -9.648 c
213.335778 -9.648 211.176 -7.488222 211.176 -4.824 c
211.176 -2.159778 213.335778 0 216 0 c
218.664222 0 220.824 -2.159778 220.824 -4.824 c
S
Q
}%
    \graphtemp=.5ex
    \advance\graphtemp by 0.067in
    \rlap{\kern 3.000in\lower\graphtemp\hbox to 0pt{\hss \scriptsize $n$\hss}}%
\pdfliteral{
q [] 0 d 1 J 1 j
0.576 w
0.072 w
q 0 g
203.976 -3.024 m
211.176 -4.824 l
203.976 -6.624 l
203.976 -3.024 l
B Q
0.576 w
q [0 3.6] 0 d
124.776 -4.824 m
203.976 -4.824 l
S Q
0.072 w
q 0 g
301.392 -70.92 m
307.512 -75.096 l
300.168 -74.304 l
301.392 -70.92 l
B Q
0.576 w
q [3.6 3.78566] 0 d
124.488 -6.48 m
300.744 -72.576 l
S Q
268.848 -4.824 m
268.848 -7.488222 266.688222 -9.648 264.024 -9.648 c
261.359778 -9.648 259.2 -7.488222 259.2 -4.824 c
259.2 -2.159778 261.359778 0 264.024 0 c
266.688222 0 268.848 -2.159778 268.848 -4.824 c
S
Q
}%
    \graphtemp=.5ex
    \advance\graphtemp by 0.067in
    \rlap{\kern 3.667in\lower\graphtemp\hbox to 0pt{\hss \tiny W\hss}}%
\pdfliteral{
q [] 0 d 1 J 1 j
0.576 w
0.072 w
q 0 g
252 -3.024 m
259.2 -4.824 l
252 -6.624 l
252 -3.024 l
B Q
0.576 w
220.824 -4.824 m
252 -4.824 l
S
Q
}%
    \graphtemp=\baselineskip
    \multiply\graphtemp by -1
    \divide\graphtemp by 2
    \advance\graphtemp by .5ex
    \advance\graphtemp by 0.067in
    \rlap{\kern 3.333in\lower\graphtemp\hbox to 0pt{\hss $a_n$\hss}}%
\pdfliteral{
q [] 0 d 1 J 1 j
0.576 w
0.072 w
q 0 g
303.48 -68.184 m
308.16 -73.944 l
301.32 -71.064 l
303.48 -68.184 l
B Q
0.576 w
q [3.6 3.515657] 0 d
219.816 -7.704 m
302.4 -69.624 l
S Q
316.8 -4.824 m
316.8 -7.488222 314.640222 -9.648 311.976 -9.648 c
309.311778 -9.648 307.152 -7.488222 307.152 -4.824 c
307.152 -2.159778 309.311778 0 311.976 0 c
314.640222 0 316.8 -2.159778 316.8 -4.824 c
S
Q
}%
    \graphtemp=.5ex
    \advance\graphtemp by 0.067in
    \rlap{\kern 4.333in\lower\graphtemp\hbox to 0pt{\hss \scriptsize E\hss}}%
\pdfliteral{
q [] 0 d 1 J 1 j
0.576 w
0.072 w
q 0 g
300.024 -3.024 m
307.224 -4.824 l
300.024 -6.624 l
300.024 -3.024 l
B Q
0.576 w
268.776 -4.824 m
300.024 -4.824 l
S
Q
}%
    \graphtemp=\baselineskip
    \multiply\graphtemp by -1
    \divide\graphtemp by 2
    \advance\graphtemp by .5ex
    \advance\graphtemp by 0.067in
    \rlap{\kern 4.000in\lower\graphtemp\hbox to 0pt{\hss $w$\hss}}%
\pdfliteral{
q [] 0 d 1 J 1 j
0.576 w
0.072 w
q 0 g
306.864 -65.808 m
309.312 -72.792 l
303.84 -67.824 l
306.864 -65.808 l
B Q
0.576 w
q [3.6 3.748053] 0 d
266.688 -8.784 m
305.352 -66.816 l
S Q
0.072 w
q 0 g
313.776 -64.8 m
311.976 -72 l
310.176 -64.8 l
313.776 -64.8 l
B Q
0.576 w
q [3.6 3.774857] 0 d
311.976 -9.576 m
311.976 -64.8 l
S Q
Q
}%
    \hbox{\vrule depth1.160in width0pt height 0pt}%
    \kern 4.708in
  }%
}%

\centerline{\box\graph}\\[2ex]
with $\out(T):=\ina(A)\uplus\{w\}$, $\ina(T):=\out(A)$ and $\inta(T):=\emptyset$.
To make sure that $T$ is an I/O automaton, the dashed arrows are labelled with all input actions of
$T$, except for $a_i$ (if $a_i \in \ina(T)$) for the dashed arrow departing from state $i$.
By construction, $T$ is strongly compatible with $A$ and $B$.
Now $C$ \textbf{may} $T$ (for any $C$) iff $\sigma\in\traces(C)$.
Hence $A$ \textbf{may} $T$, and thus $B$ \textbf{may} $T$, and therefore $\sigma\in\traces(B)$.
\qed
\end{proof}

\section{Must testing based on progress}\label{sec:must progress}

\begin{definitionR}{complementary}
An I/O automaton $T$ is \emph{complementary} to I/O automaton $A$ if $\out(T)=\ina(A) \uplus \{w\}$,
$\ina(T)=\out(A)$ and $\inta(T) \cap \inta(A) = \emptyset$.
\end{definitionR}
In this case $T$ and $A$ are also strongly compatible, so that $T\|A$ is defined, and $\ina(T\|A)=\emptyset$.
I now show that for the definition of $\sqsubseteq_{\rm must}^{\it Pr}$ it makes no difference
whether one restricts the tests $T$ that may be used to compare two I/O automata $A$ and $B$
to ones that are complementary to $A$ and $B$.

For use in the following proof,
define the relation $\equiv$ between I/O automata by $C \equiv D$ iff
$\states(C)=\states(D) \wedge \start(C)=\start(D) \wedge \steps(C)=\steps(D)$.
Note that $T\|A \equiv T'\|A$ implies that $A$ \textbf{must} $T$ iff $A$ \textbf{must} $T'$.
\begin{propositionR}{complementary}
$A \sqsubseteq_{\rm must}^{\it Pr} B$ iff
$\ina(A)\mathbin=\ina(B) \wedge \out(A)\mathbin=\out(B)$ and
$A$ \textbf{must} $T$ implies $B$ \textbf{must} $T$ for each test $T$ that is complementary to $A$ and $B$.
\end{propositionR}

\begin{proof}
Suppose $A \sqsubseteq_{\rm must}^{\it Pr} B$. Then $\ina(A)\mathbin=\ina(B) \wedge \out(A)\mathbin=\out(B)$ and
\mbox{$A$ \textbf{must} $T$} implies $B$ \textbf{must} $T$ for each test $T$ that is strongly compatible
with $A$ and $B$, and thus certainly for each test $T$ that is complementary to $A$ and $B$.

Now suppose $\ina(A)\mathbin=\ina(B) \wedge \out(A)\mathbin=\out(B)$ but $A \not\sqsubseteq_{\rm must}^{\it Pr} B$.
Then there is a test $T$, strongly compatible with $A$ and $B$, such that $A$ \textbf{must} $T$, yet
$\neg (B~{\bf must}~T)$. It suffices to find a test $T''$ with the same properties that is
moreover  complementary to $A$ and $B$.

First modify $T$ into $T'$ by adding $\ext(A)\setminus\ext(T)$ to $\ina(T')$,
while adding a loop $(s,a,s)$ to $\steps(T')$ for each state $s \in \states(T')$
and each $a \in \ext(A)\setminus\ext(T)$.
Now $T\|A = T'\|A$ and $T\|B = T'\|B$,
and thus $A$ \textbf{must} $T'$, yet $\neg (B~{\bf must}~T')$. Moreover,
$\ext(A)=\ext(B)\subseteq \ext(T')$.

Modify $T'$ further into $T''$ by reclassifying any action $a \in \ina(T') \cap \ina(A)$ as an output action of $T''$
and any $a \in \ext(T')\setminus(\ext(A)\uplus\{w\})$ as an internal action of $T''$.
How $\pa(T'')$ is defined is immaterial. Then $T'\|A \equiv T''\|A$ and $T'\|B \equiv T''\|B$, and thus
$A$ \textbf{must} $T''$, yet $\neg (B~{\bf must}~T'')$.
Now $\out(T'')\mathbin=\ina(A)\uplus\{w\}$, $\ina(T'')\mathbin=\out(A)$, $\inta(T'') \cap \inta(A) = \emptyset$ and
$\inta(T'') \cap \inta(B) = \emptyset$.
\qed
\end{proof}

\noindent
Using the characterisation of \pr{complementary} as definition, the preorder $\sqsubseteq_{\rm must}^{\it Pr}$
on I/O automata has been studied by Segala \cite[Section~7]{Se97}. There it was related to the
\emph{quiescent trace} preorder $\sqsubseteq_Q$ defined by Vaandrager \cite{Va91}.
Similar as for the preorders of \sect{IO}, I write $S \sqsubseteq_Q I$ for what was denoted
$I \sqsubseteq_Q S$ in \cite{Se97}, and $I \sqsubseteq_{\it qT} S$ in \cite{Va91}.

\begin{definitionR}{quiescent}
An execution $\alpha$ is \emph{quiescent} if it is finite and its last state enables only input actions.
Let $\qtraces(A) \mathbin{:=} \{\trace(\alpha)\! \mid \mbox{$\alpha$ is a quiescent execution of $A$}\}$. Now\vspace{-1ex}
\[S \sqsubseteq_Q I :\Leftrightarrow S \sqsubseteq_T I \wedge \qtraces(I) \subseteq \qtraces(S)\makebox[0pt][l]{\,.}\]
\end{definitionR}
An I/O automaton is \emph{finitely branching} iff each of its states enables finitely many transitions;
it is \emph{strongly convergent} if it has no infinite execution $\alpha$ with $\trace(\alpha)$ finite,
i.e., no execution with an infinite suffix of only internal actions.

\begin{theoremRc}{must quiescent}{{\cite[Thm.~7.3]{Se97}}}
Let $A$ and $B$ be finitely branching and strongly convergent I/O automata. Then
                $A  \sqsubseteq_{\rm must}^{\it Pr} B$ iff $A \sqsubseteq_Q B$.
\end{theoremRc}
Note that an execution is quiescent iff it is fair and finite.
By \cite[Thm.~5.7]{Se97}, if $A$ is strongly convergent then $A \sqsubseteq_F B$ implies $A  \sqsubseteq_Q B$.
(For let $A \sqsubseteq_F B$. If $\sigma\in \qtraces(B)$, then $\sigma\in\fairtraces(B) \subseteq \fairtraces(A)$
so $A$ has a fair execution $\alpha$ with $\trace(\alpha)=\sigma$. As $A$ is strongly convergent,
$\alpha$ is finite. Hence $\sigma\in \qtraces(A)$.)
This does not hold when dropping the side condition of strong convergence. Take $A = \!$
\expandafter\ifx\csname graph\endcsname\relax
   \csname newbox\expandafter\endcsname\csname graph\endcsname
\fi
\ifx\graphtemp\undefined
  \csname newdimen\endcsname\graphtemp
\fi
\expandafter\setbox\csname graph\endcsname
 =\vtop{\vskip 0pt\hbox{%
\pdfliteral{
q [] 0 d 1 J 1 j
0.576 w
0.576 w
16.992 -2.808 m
16.992 -4.358816 15.734816 -5.616 14.184 -5.616 c
12.633184 -5.616 11.376 -4.358816 11.376 -2.808 c
11.376 -1.257184 12.633184 0 14.184 0 c
15.734816 0 16.992 -1.257184 16.992 -2.808 c
S
0.072 w
q 0 g
4.104 -1.008 m
11.304 -2.808 l
4.104 -4.608 l
4.104 -1.008 l
B Q
0.576 w
0 -2.808 m
4.104 -2.808 l
S
Q
}%
    \hbox{\vrule depth0.079in width0pt height 0pt}%
    \kern 0.236in
  }%
}%

\raisebox{6pt}[0pt][0pt]{\box\graph}
and $B = \!$
\expandafter\ifx\csname graph\endcsname\relax
   \csname newbox\expandafter\endcsname\csname graph\endcsname
\fi
\ifx\graphtemp\undefined
  \csname newdimen\endcsname\graphtemp
\fi
\expandafter\setbox\csname graph\endcsname
 =\vtop{\vskip 0pt\hbox{%
\pdfliteral{
q [] 0 d 1 J 1 j
0.576 w
0.576 w
16.992 -3.96 m
16.992 -5.510816 15.734816 -6.768 14.184 -6.768 c
12.633184 -6.768 11.376 -5.510816 11.376 -3.96 c
11.376 -2.409184 12.633184 -1.152 14.184 -1.152 c
15.734816 -1.152 16.992 -2.409184 16.992 -3.96 c
S
0.072 w
q 0 g
4.104 -2.16 m
11.304 -3.96 l
4.104 -5.76 l
4.104 -2.16 l
B Q
0.576 w
0 -3.96 m
4.104 -3.96 l
S
0.072 w
q 0 g
23.544 -1.296 m
16.2 -1.944 l
22.392 2.016 l
23.544 -1.296 l
B Q
0.576 w
16.2 -5.976 m
19.44 -7.092 l
21.6432 -7.85088 24.03936 -8.208 26.928 -8.208 c
29.81664 -8.208 31.176 -6.86016 31.176 -3.996 c
31.176 -1.13184 29.81664 0.216 26.928 0.216 c
24.03936 0.216 21.74688 -0.09504 19.764 -0.756 c
16.848 -1.728 l
S
Q
}%
    \graphtemp=.5ex
    \advance\graphtemp by 0.055in
    \rlap{\kern 0.492in\lower\graphtemp\hbox to 0pt{\hss $\tau$\hss}}%
    \hbox{\vrule depth0.110in width0pt height 0pt}%
    \kern 0.492in
  }%
}%

\raisebox{7pt}[0pt][0pt]{\box\graph}\hspace{5pt}
with $\acts(A)=\emptyset$ and $\acts(B)=\inta(B)=\{\tau\}$.
Then $A \equiv_F B$, yet $A \not\sqsubseteq_Q B$ (and $A \not\sqsubseteq_{\rm must}^{\it Pr} B$).

Even restricted to finitely branching and strongly convergent I/O automata, 
$A \mathbin\sqsubseteq_Q B$ does not imply $A \mathbin\sqsubseteq_F B$. This is illustrated by
\cite[Examples 5.1 and~5.2]{Se97}.

\section{Must testing based on fairness}\label{sec:fair must testing}

As explained in \sect{testing}, the notion of must testing is naturally parametrised
by the choice of a completeness criterion. As I/O automata are already equipped with
a completeness criteria, namely the notion of fairness from \df{fairness},
the most appropriate form of must testing for I/O automata takes this concept of fairness
as its parameter, rather than the default completeness criterion used in \sect{must progress}.

A problem in properly defining a must-testing preorder $\sqsubseteq_{\rm must}^F$ involves the
definition of the operator $[\;\;]$ employed in \df{testing}. In the context of standard automata,
this operator reclassifies all its external actions, except for the success action $w$, as internal.
When applied to I/O automaton $A$, it is not a priori clear how to define $\pa([A])$, for this is a
partition of the set of locally-controlled actions into tasks, and when changing an input action
into a locally-controlled action, one lacks guidance on which task to allocate it to.
This was a not a problem in \sect{must progress}, as there the must-testing preorder 
$\sqsubseteq_{\rm must}^{\it Pr}$ depends in no way on $\pa$.

Below I inventorise various solutions to this problem, which gives rise to three possible
definitions of $\sqsubseteq_{\rm must}^F$. Then I show in \sect{agrees} that all three resulting
preorders coincide, so that it doesn't matter on which of the definitions one settles. Moreover,
these preorders all turn out to coincide with the fair preorder $\sqsubseteq_F$ that comes with
I/O automata.

My first (and default) solution is to simply drop the operator $[\;\;]$ from \df{testing}:

\begin{definitionR}{fair testing}
An I/O automaton $A$ \emph{must pass} a test $T$ \emph{fairly}---$A~\textbf{must}^F \,T$---if
each fair execution of $T\|A$ is successful.
Write $A \sqsubseteq_{\rm must}^F B$ if  $\ina(A)\mathbin=\ina(B) \wedge \out(A)\mathbin=\out(B)$ and
$A~\textbf{must}^F\,T$ implies $B~\textbf{must}^F\,T$ for each test $T$ that is strongly compatible with $A$ and $B$.
\end{definitionR}
This is a plausible approach, as none of the testing preorders discussed in
Sections~\ref{sec:testing}--\ref{sec:must progress} would change at all were the
operator $[\;\;]$ dropped from \df{testing}. This is the case because the set of executions,
successful executions and complete executions of an automaton $A$ is independent of the status (input,
output or internal) of the actions of $A$.

The above begs the question why I bothered to employ the operator $[\;\;]$ in \df{testing} in the
first place. The main reason is that the theory of testing \cite{DH84} was developed in the context
of CCS, where each synchronisation of an action from a test with one from a tested process
yields an internal action $\tau$. \df{testing} recreates this theory using the operator $\|$ from CSP
\cite{Ho85} and I/O automata \cite{LT89}, but as here synchronised actions are not internal, they
have to be made internal to obtain the same effect. A second reason concerns the argument used
towards the end of \sect{testing} for not parametrising notions of testing with a set $B$ of actions
that can be blocked; this argument hinges on all relevant actions being internal.

My second solution is to restrict the set of allowed tests $T$ for comparing I/O automata $A$ and $B$
to those for which $\ina(T\|A)=\ina(T\|B)=\emptyset$. This is the case iff $\ina(T) \subseteq \out(A)$
and $\ina(A) \subseteq \out(T)$. In that case $[T\|A]$ and $[T\|B]$ are
trivial to define, as the set of locally-controlled actions stays the same. Moreover, it makes no
difference whether this operator is included in the definition of {\bf must} or not, as the set of
fair executions of a process is not affected by a reclassification of output actions as internal actions.

\begin{definitionR}{fair testing empty}
Write \(A \mathrel{\raisebox{6pt}{$\scriptscriptstyle\emptyset$}\!\!\sqsubseteq_{\rm must}^F} B\)
if $\ina(A)\mathbin=\ina(B) \wedge \out(A)\mathbin=\out(B)$ and moreover
$A~\textbf{must}^F\,T$ implies $B~\textbf{must}^F\,T$ for each test $T$ that is strongly compatible
with $A$ and $B$, and for which $\ina(T\|A)=\ina(T\|B)=\emptyset$.
\end{definitionR}
A small variation of this idea restricts the set of allowed tests even further, namely to the ones
that are complementary to $A$ and $B$, as defined in \df{complementary}. This yield a fair version of
the must-testing preorder employed in \cite{Se97}.

\begin{definitionR}{fair testing complementary}
Write \(A \mathrel{\raisebox{6.5pt}{$\scriptscriptstyle{\rm cm}$}\!\!\!\!\sqsubseteq_{\rm must}^F} B\)
if $\ina(A)\mathbin=\ina(B) \wedge \out(A)\mathbin=\out(B)$ and
$A~\textbf{must}^F\,T$ implies $B~\textbf{must}^F\,T$ for each $T$ that is complementary
to $A$ and $B$.
\end{definitionR}

\noindent
As a last solution I consider tests $T$ that are not restricted as in Defs.~\ref{df:fair testing empty} or
\ref{df:fair testing complementary}, while looking for elegant ways to define $[T\|A]$ and $[T\|B]$.
First of all, note that no generality is lost when restricting to tests $T$ such that
$\ext(A) (=\ext(B)) \subseteq\ext(T)$, regardless how the operator $[\;\;]$ is defined.
Namely, employing the first conversion from the proof of \pr{complementary},
any test $T$ that is strongly compatible with I/O automata $A$ and $B$ can converted
into a test $T'$ satisfying this requirement, and such that $T\|A = T'\|A$ and $T\|B = T'\|B$.

An application of $[\;\;]$ to $T\|A$ consists of reclassifying external actions of $T\|A$ as internal actions.
However, since for the definition of the testing preorders it makes no difference whether an
action in $T\|A$ is an internal or an output action, one can just as well use an operator $[\;\;]'$ that
merely reclassifies input actions of $T\|A$ as output actions. Note that
$\ina(T\|A) \subseteq \ina(T)$, using that $\ext(A) \subseteq\ext(T)$.
Let $T^*$ be a result of adapting the test $T$ by reclassifying the actions in $\ina(T\|A)$
from input actions of $T$ into output actions of $T$; the test $T^*$ is not uniquely defined, as
there are various ways to fill in $\pa(T^*)$.
\begin{observationR}{reclass}
Apart from the problematic definition of $\pa([T\|A]')$,
the I/O automaton $[T\|A]'$ is the very same as $T^*\|A$.
\end{observationR}
In other words, the reclassification of input into output actions can just as well be done on the
test, instead of on the composition of test and tested automaton.
The advantage of this approach is that the problematic definition of $\pa([T\|A]')$ is moved
to the test as well.
Now one can use $T^*\|A$ instead of $[T\|A]'$ in the definition of must testing
for any desired definition of $\pa(T^*)$.
This amounts to choosing any test $T^*$ with $\ina(T^*\|A)=\emptyset$.
It makes this solution equivalent to the one of \df{fair testing empty}.

\section{Action-based must testing}\label{sec:action-based}

The theory of testing from \cite{DH84} employs the success action $w$ merely to mark success states;
an execution is successful iff it contains a state in which $w$ is enabled. In \cite{DGHM08} this is
dubbed \emph{state-based testing}. Segala~\cite{Se96} (in a setting with probabilistic automata)
uses another mode of testing, called \emph{action-based} in \cite{DGHM08}, in which an execution is
defined to be successful iff it contains the action $w$.

Although the state-based and action-based may-testing preorders obviously coincide,
the state-based and action-based must-testing preorders do not, at least when employing the default
completeness criterion. An example showing the difference is given in \cite{DGHM08}.
It involves two automata $A$ and $B$, which can in fact be seen as I/O automata, such that
$A \not\sqsubseteq_{\rm must}^{\it Pr} B$, yet
\(A \mathrel{\raisebox{6pt}{$\scriptscriptstyle{\rm ab}$}\!\!\!\!\equiv_{\rm must}^{\it Pr}} B\).
Here \(\raisebox{6pt}{$\scriptscriptstyle{\rm ab}$}\!\!\!\!\equiv_{\rm must}^{\it Pr}\)
is the action-based version of $\equiv_{\rm must}^{\it Pr}$.

So far I have considered only state-based testing preorders on I/O automata.
Let  \raisebox{0pt}[0pt]{\(\raisebox{6.5pt}{$\scriptscriptstyle{\rm ab}$}\!\!\!\!\sqsubseteq_{\rm must}^{\it F}\)}
be the action-based version of $\sqsubseteq_{\rm must}^{\it F}$. It is defined as in \df{fair testing},
but using $\textbf{must}^F_{\rm ab}$ instead of \textbf{must}$^F$. Here $A$ \textbf{must}$^F_{\rm ab}$ $T$
holds iff each fair trace of $T\|A$ contains the action $w$. Below I will show that when taking the
notion of fairness from \cite{LT89} as completeness criterion, state-based and action-based must
testing yields the same result, i.e.,
\raisebox{0pt}[0pt]{\(\raisebox{6.5pt}{$\scriptscriptstyle{\rm ab}$}\!\!\!\!\sqsubseteq_{\rm must}^{\it F}\)}
equals  $\sqsubseteq_{\rm must}^{\it F}$. In fact, I need this result in my proof that  $\sqsubseteq_{\rm must}^{\it F}$
coincides with  $\sqsubseteq_{\it F}$.

\section{Fair must testing agrees with the fair traces preorder}\label{sec:agrees}

The following theorem states that the must-testing preorder on I/O automata based on the
completeness criterion of fairness that is native to I/O automata, in each of the four forms
discussed in Sections~\ref{sec:fair must testing} and \ref{sec:action-based}, coincides with the
standard preorder of I/O automata based on reverse inclusion of fair traces.

\begin{theoremR}{fair must testing}
  $A \mathop{\raisebox{6.5pt}{$\scriptscriptstyle{\rm ab}$}\!\!\!\!\sqsubseteq_{\rm must}^{\it F}} B$
  iff
  $A  \mathop\sqsubseteq_{\rm must}^{\it F} B$
  iff
  \(A \mathop{\raisebox{6pt}{$\scriptscriptstyle\emptyset$}\!\!\sqsubseteq_{\rm must}^F} B\)
  iff
  \(A \mathop{\raisebox{6.5pt}{$\scriptscriptstyle{\rm cm}$}\!\!\!\!\sqsubseteq_{\rm must}^F} B\)
  iff
  $A \mathop{\sqsubseteq_F} B$.
\end{theoremR}

\begin{proof}
Suppose $A \mathbin\sqsubseteq_F B$, i.e., $\ina(A)\mathbin=\ina(B) \wedge \out(A)\mathbin=\out(B)$
and $\fairtraces(B) \mathbin\subseteq \fairtraces(A)$, and let $T$ be any test that is strongly
compatible with $A$ and $B$.
Since $\sqsubseteq_F$ is a precongruence for composition (cf. \sect{IO}),
$\fairtraces(T\|B) \subseteq \traces(T\|A)$.
Since for action-based must testing $C$ \textbf{must}$^F_{\rm ab}$ $T$ (for any $C$) iff $w$ occurs in
each fair trace $\sigma \in \fairtraces(T\|C)$, 
it follows that $A~\textbf{must}^F_{\rm ab}~T$ implies $B~\textbf{must}^F_{\rm ab}~T$.
Thus \raisebox{0pt}[0pt]{$A \mathop{\raisebox{6.5pt}{$\scriptscriptstyle{\rm ab}$}\!\!\!\!\sqsubseteq_{\rm must}^{\it F}} B$}.

Now suppose \raisebox{0pt}[0pt]{$A \mathop{\raisebox{6.5pt}{$\scriptscriptstyle{\rm ab}$}\!\!\!\!\sqsubseteq_{\rm must}^{\it F}} B$}.
In order to show that $A \sqsubseteq_{\rm must}^{\it F} B$, suppose that $A~\textbf{must}^F~T$, where
$T$ is a test that is strongly compatible with $A$ and $B$. Let the test $T^*$ be obtained from $T$
by (i) dropping all transitions $(s,a,s')\in\steps(T)$ for $s$ a success state and $a \neq w$, and
(ii) adding a loop $(s,a,s)$ for each success state $s$ and $a \in \ina(T)$. Since for state-based
must testing it is irrelevant what happens after encountering  a success state, one has
\begin{equation}\label{slimming tests}
  C~\textbf{must}^F~T ~~\mbox{iff}~~ C~\textbf{must}^F~T^*
\end{equation}
for each I/O automaton $C$. Moreover, I claim that for each $C$ one has
\begin{equation}\label{action vs state}
  C~\textbf{must}^F~T^* ~~\mbox{iff}~~ C~\textbf{must}^F_{\rm ab}~T^*.
\end{equation}
Here ``if'' is trivial. For ``only if'', let $\alpha$ be a fair execution of $T^*\|C$,
and suppose, towards a contradiction, that $\alpha$ contains a success state $(s,r)$,
with $s$ a success state of $T^*$ and $r$ a state of $C$,
but does not contain the success action $w$. Let $\alpha'$ be the suffix of $\alpha$ starting with
the first occurrence of $(s,r)$. Then all states of $\alpha'$ have the form $(s,r')$,
and the action $w$ is enabled in each of these states. Let $\C \in \pa(T^*\|C)$ be the task containing $w$.
Since $w$ is a locally controlled action of $T^*$, by \df{composition} all members of $\C$ must be
locally controlled actions of $T^*$. No such action can occur in $\alpha'$. This contradicts the
assumption that $\alpha$ is fair (cf.\ \df{fairness}), and thereby concludes the proof of (\ref{action vs state}).

From the assumption $A~\textbf{must}^F~T$ one obtains $A~\textbf{must}^F_{\rm ab}~T^*$ by (\ref{slimming tests})
and (\ref{action vs state}), and $B~\textbf{must}^F_{\rm ab}~T^*$ by the assumption that
\raisebox{0pt}[0pt]{$A \mathop{\raisebox{6.5pt}{$\scriptscriptstyle{\rm ab}$}\!\!\!\!\sqsubseteq_{\rm must}^{\it F}} B$}.
Hence $B~\textbf{must}^F~T$ by (\ref{action vs state}) and (\ref{slimming tests}). Thus $A \sqsubseteq_{\rm must}^{\it F} B$.

That $A  \mathop\sqsubseteq_{\rm must}^{\it F} B$ implies
\(A \mathop{\raisebox{6pt}{$\scriptscriptstyle\emptyset$}\!\!\sqsubseteq_{\rm must}^F} B\)
is trivial.

That \(A \mathop{\raisebox{6pt}{$\scriptscriptstyle\emptyset$}\!\!\sqsubseteq_{\rm must}^F} B\)
implies \(A \mathop{\raisebox{6.5pt}{$\scriptscriptstyle{\rm cm}$}\!\!\!\!\sqsubseteq_{\rm must}^F} B\)
is also trivial.

Finally, suppose \(A \mathop{\raisebox{6.5pt}{$\scriptscriptstyle{\rm cm}$}\!\!\!\!\sqsubseteq_{\rm must}^F} B\).
Then $\ina(A)\mathbin=\ina(B) \wedge \out(A)\mathbin=\out(B)$.
Let $\sigma=a_1 a_2 \dots a_n \in \fairtraces(B)$. Let $T$ be the test automaton\\
\expandafter\ifx\csname graph\endcsname\relax
   \csname newbox\expandafter\endcsname\csname graph\endcsname
\fi
\ifx\graphtemp\undefined
  \csname newdimen\endcsname\graphtemp
\fi
\expandafter\setbox\csname graph\endcsname
 =\vtop{\vskip 0pt\hbox{%
\pdfliteral{
q [] 0 d 1 J 1 j
0.576 w
0.576 w
268.848 -76.824 m
268.848 -79.488222 266.688222 -81.648 264.024 -81.648 c
261.359778 -81.648 259.2 -79.488222 259.2 -76.824 c
259.2 -74.159778 261.359778 -72 264.024 -72 c
266.688222 -72 268.848 -74.159778 268.848 -76.824 c
S
Q
}%
    \graphtemp=.5ex
    \advance\graphtemp by 1.067in
    \rlap{\kern 3.667in\lower\graphtemp\hbox to 0pt{\hss \tiny W\hss}}%
\pdfliteral{
q [] 0 d 1 J 1 j
0.576 w
316.8 -76.824 m
316.8 -79.488222 314.640222 -81.648 311.976 -81.648 c
309.311778 -81.648 307.152 -79.488222 307.152 -76.824 c
307.152 -74.159778 309.311778 -72 311.976 -72 c
314.640222 -72 316.8 -74.159778 316.8 -76.824 c
S
Q
}%
    \graphtemp=.5ex
    \advance\graphtemp by 1.067in
    \rlap{\kern 4.333in\lower\graphtemp\hbox to 0pt{\hss \scriptsize E\hss}}%
\pdfliteral{
q [] 0 d 1 J 1 j
0.576 w
0.072 w
q 0 g
300.024 -75.024 m
307.224 -76.824 l
300.024 -78.624 l
300.024 -75.024 l
B Q
0.576 w
268.776 -76.824 m
300.024 -76.824 l
S
Q
}%
    \graphtemp=\baselineskip
    \multiply\graphtemp by -1
    \divide\graphtemp by 2
    \advance\graphtemp by .5ex
    \advance\graphtemp by 1.067in
    \rlap{\kern 4.000in\lower\graphtemp\hbox to 0pt{\hss $w$\hss}}%
\pdfliteral{
q [] 0 d 1 J 1 j
0.576 w
0.072 w
q 0 g
322.776 -72.72 m
315.36 -73.44 l
321.624 -69.336 l
322.776 -72.72 l
B Q
0.576 w
q [3.6] 0 d
315.36 -80.208 m
320.868 -82.116 l
324.61344 -83.41344 328.68 -84.024 333.576 -84.024 c
338.472 -84.024 340.776 -81.72 340.776 -76.824 c
340.776 -71.928 338.472 -69.624 333.576 -69.624 c
328.68 -69.624 324.72864 -70.18848 321.228 -71.388 c
316.08 -73.152 l
S
Q
28.8 -4.824 m
28.8 -7.488222 26.640222 -9.648 23.976 -9.648 c
21.311778 -9.648 19.152 -7.488222 19.152 -4.824 c
19.152 -2.159778 21.311778 0 23.976 0 c
26.640222 0 28.8 -2.159778 28.8 -4.824 c
S
Q
}%
    \graphtemp=.5ex
    \advance\graphtemp by 0.067in
    \rlap{\kern 0.333in\lower\graphtemp\hbox to 0pt{\hss \scriptsize 1\hss}}%
\pdfliteral{
q [] 0 d 1 J 1 j
0.576 w
0.072 w
q 0 g
12.024 -3.024 m
19.224 -4.824 l
12.024 -6.624 l
12.024 -3.024 l
B Q
0.576 w
0 -4.824 m
12.024 -4.824 l
S
0.072 w
q 0 g
253.008 -71.64 m
259.416 -75.456 l
252 -75.096 l
253.008 -71.64 l
B Q
0.576 w
q [3.6 3.593104] 0 d
28.584 -6.192 m
252.504 -73.368 l
S Q
0.072 w
q 0 g
252.072 -74.736 m
259.2 -76.824 l
251.928 -78.336 l
252.072 -74.736 l
B Q
0.576 w
251.988273 -76.626718 m
172.613398 -72.798626 95.430561 -49.324402 27.365337 -8.310376 c
S
Q
}%
    \graphtemp=.5ex
    \advance\graphtemp by 0.800in
    \rlap{\kern 1.667in\lower\graphtemp\hbox to 0pt{\hss $\tau$\hss}}%
\pdfliteral{
q [] 0 d 1 J 1 j
0.576 w
76.824 -4.824 m
76.824 -7.488222 74.664222 -9.648 72 -9.648 c
69.335778 -9.648 67.176 -7.488222 67.176 -4.824 c
67.176 -2.159778 69.335778 0 72 0 c
74.664222 0 76.824 -2.159778 76.824 -4.824 c
S
Q
}%
    \graphtemp=.5ex
    \advance\graphtemp by 0.067in
    \rlap{\kern 1.000in\lower\graphtemp\hbox to 0pt{\hss \scriptsize 2\hss}}%
\pdfliteral{
q [] 0 d 1 J 1 j
0.576 w
0.072 w
q 0 g
59.976 -3.024 m
67.176 -4.824 l
59.976 -6.624 l
59.976 -3.024 l
B Q
0.576 w
28.8 -4.824 m
59.976 -4.824 l
S
Q
}%
    \graphtemp=\baselineskip
    \multiply\graphtemp by -1
    \divide\graphtemp by 2
    \advance\graphtemp by .5ex
    \advance\graphtemp by 0.067in
    \rlap{\kern 0.667in\lower\graphtemp\hbox to 0pt{\hss $a_1$\hss}}%
\pdfliteral{
q [] 0 d 1 J 1 j
0.576 w
0.072 w
q 0 g
253.368 -70.92 m
259.488 -75.096 l
252.144 -74.304 l
253.368 -70.92 l
B Q
0.576 w
q [3.6 3.788356] 0 d
76.464 -6.48 m
252.792 -72.576 l
S Q
0.072 w
q 0 g
252.36 -73.872 m
259.2 -76.824 l
251.784 -77.472 l
252.36 -73.872 l
B Q
0.576 w
252.114324 -75.735994 m
189.320757 -65.18829 129.253105 -42.253206 75.40913 -8.26626 c
S
Q
}%
    \graphtemp=.5ex
    \advance\graphtemp by 0.700in
    \rlap{\kern 2.000in\lower\graphtemp\hbox to 0pt{\hss $\tau$\hss}}%
\pdfliteral{
q [] 0 d 1 J 1 j
0.576 w
124.848 -4.824 m
124.848 -7.488222 122.688222 -9.648 120.024 -9.648 c
117.359778 -9.648 115.2 -7.488222 115.2 -4.824 c
115.2 -2.159778 117.359778 0 120.024 0 c
122.688222 0 124.848 -2.159778 124.848 -4.824 c
S
Q
}%
    \graphtemp=.5ex
    \advance\graphtemp by 0.067in
    \rlap{\kern 1.667in\lower\graphtemp\hbox to 0pt{\hss \scriptsize 3\hss}}%
\pdfliteral{
q [] 0 d 1 J 1 j
0.576 w
0.072 w
q 0 g
108 -3.024 m
115.2 -4.824 l
108 -6.624 l
108 -3.024 l
B Q
0.576 w
76.824 -4.824 m
108 -4.824 l
S
Q
}%
    \graphtemp=\baselineskip
    \multiply\graphtemp by -1
    \divide\graphtemp by 2
    \advance\graphtemp by .5ex
    \advance\graphtemp by 0.067in
    \rlap{\kern 1.333in\lower\graphtemp\hbox to 0pt{\hss $a_2$\hss}}%
\pdfliteral{
q [] 0 d 1 J 1 j
0.576 w
0.072 w
q 0 g
254.088 -69.84 m
259.704 -74.664 l
252.432 -73.008 l
254.088 -69.84 l
B Q
0.576 w
q [3.6 3.802801] 0 d
124.272 -6.912 m
253.296 -71.424 l
S Q
0.072 w
q 0 g
252.936 -72.864 m
259.2 -76.824 l
251.784 -76.248 l
252.936 -72.864 l
B Q
0.576 w
252.353079 -74.658535 m
206.249764 -59.195059 162.763738 -36.816028 123.383939 -8.287687 c
S
Q
}%
    \graphtemp=.5ex
    \advance\graphtemp by 0.467in
    \rlap{\kern 2.333in\lower\graphtemp\hbox to 0pt{\hss $\tau$\hss}}%
\pdfliteral{
q [] 0 d 1 J 1 j
0.576 w
220.824 -4.824 m
220.824 -7.488222 218.664222 -9.648 216 -9.648 c
213.335778 -9.648 211.176 -7.488222 211.176 -4.824 c
211.176 -2.159778 213.335778 0 216 0 c
218.664222 0 220.824 -2.159778 220.824 -4.824 c
S
Q
}%
    \graphtemp=.5ex
    \advance\graphtemp by 0.067in
    \rlap{\kern 3.000in\lower\graphtemp\hbox to 0pt{\hss \scriptsize $n$\hss}}%
\pdfliteral{
q [] 0 d 1 J 1 j
0.576 w
0.072 w
q 0 g
203.976 -3.024 m
211.176 -4.824 l
203.976 -6.624 l
203.976 -3.024 l
B Q
0.576 w
q [0 3.6] 0 d
124.776 -4.824 m
203.976 -4.824 l
S Q
0.072 w
q 0 g
258.84 -65.808 m
261.36 -72.792 l
255.816 -67.824 l
258.84 -65.808 l
B Q
0.576 w
q [3.6 3.748053] 0 d
218.664 -8.784 m
257.328 -66.816 l
S Q
0.072 w
q 0 g
257.472 -66.672 m
260.64 -73.44 l
254.664 -68.976 l
257.472 -66.672 l
B Q
0.576 w
256.063493 -67.891763 m
241.29772 -49.493334 227.908325 -30.031435 216.004155 -9.66428 c
S
Q
}%
    \graphtemp=.5ex
    \advance\graphtemp by 0.467in
    \rlap{\kern 3.133in\lower\graphtemp\hbox to 0pt{\hss $\tau$\hss}}%
\pdfliteral{
q [] 0 d 1 J 1 j
0.576 w
268.848 -4.824 m
268.848 -7.488222 266.688222 -9.648 264.024 -9.648 c
261.359778 -9.648 259.2 -7.488222 259.2 -4.824 c
259.2 -2.159778 261.359778 0 264.024 0 c
266.688222 0 268.848 -2.159778 268.848 -4.824 c
S
Q
}%
    \graphtemp=.5ex
    \advance\graphtemp by 0.067in
    \rlap{\kern 3.667in\lower\graphtemp\hbox to 0pt{\hss \scriptsize S\hss}}%
\pdfliteral{
q [] 0 d 1 J 1 j
0.576 w
0.072 w
q 0 g
252 -3.024 m
259.2 -4.824 l
252 -6.624 l
252 -3.024 l
B Q
0.576 w
220.824 -4.824 m
252 -4.824 l
S
Q
}%
    \graphtemp=\baselineskip
    \multiply\graphtemp by -1
    \divide\graphtemp by 2
    \advance\graphtemp by .5ex
    \advance\graphtemp by 0.067in
    \rlap{\kern 3.333in\lower\graphtemp\hbox to 0pt{\hss $a_n$\hss}}%
\pdfliteral{
q [] 0 d 1 J 1 j
0.576 w
0.072 w
q 0 g
265.824 -64.8 m
264.024 -72 l
262.224 -64.8 l
265.824 -64.8 l
B Q
0.576 w
q [3.6 3.774857] 0 d
264.024 -9.576 m
264.024 -64.8 l
S Q
Q
}%
    \hbox{\vrule depth1.160in width0pt height 0pt}%
    \kern 4.708in
  }%
}%

\centerline{\box\graph}\\[2ex]
with $\out(T):=\ina(A)\uplus\{w\}$, $\ina(T):=\out(A)$ and $\inta(T):=\{\tau\}$.
The dashed arrows are labelled with all input actions of $T$,
except for $a_i$ (if $a_i \in \ina(T)$) for the dashed arrow departing from state $i$.
By construction, $T$ is complementary to $A$ and $B$.
Now $C$ \textbf{must} $T$ (for any $C$) iff $\sigma\not\in\fairtraces(C)$.
Hence $B$ \textbf{may not} $T$, and thus $A$ \textbf{may not} $T$, and therefore $\sigma\in\fairtraces(A)$.

The case that $\sigma=a_1 a_2 \dots \in \fairtraces(B)$ is infinite goes likewise, but without the
state {\scriptsize S} in $T$. Hence $A \mathop{\sqsubseteq_F} B$.
\qed
\end{proof}

\section{Reward testing}

The reward testing preorder taking the notion of fairness from \df{fairness} as underlying
completeness criterion can be defined on I/O automata by analogy of Definitions~\ref{df:fair testing},
\ref{df:fair testing empty} or~\ref{df:fair testing complementary}. Here I take the one that follows
\df{fair testing}, as it is clearly the strongest, i.e., with its kernel making the most distinctions.

\begin{definitionR}{reward testing}
Write $A \sqsubseteq_{\rm reward}^F B$ if $\ina(A)\mathbin=\ina(B) \wedge \out(A)\mathbin=\out(B)$ and
for each reward test $T$ that is strongly compatible with $A$ and $B$ and for each fair
execution $\beta$ of $T\|B$
there is a fair execution $\alpha$ of $T\|A$ with $\reward(\alpha) \mathbin\leq \reward(\beta)$.
\end{definitionR}
When taking progress as underlying completeness criterion, reward testing is stronger than must
testing; the opening page of \cite{vG19b} shows an example where reward testing makes useful
distinctions that are missed by may as well as must testing. 
When moving to fairness as the underlying completeness criterion, must testing no longer misses that
example, and in fact must testing becomes equally strong as reward testing.
In order to show this, I will use the following notation.

\begin{definitionR}{projection}
Let $A_1$ and $A_2$ be two strongly compatible I/O automata.
A state $\vec{s}$ of $A_1\|A_2$ is a pair $(\vec{s}\,[1],\vec{s}\,[2])$ with $\vec{s}\,[k]\in\states(A_k)$
for $k=1,2$.
Let $\alpha = \vec{s}_0, a_1, \vec{s}_1, a_2, \dots$ be an execution of $A_1\|A_2$.
The projection $\alpha[k]$ of $\alpha$ to the $k^{\rm th}$ component $A_k$, for $k=1,2$, is obtained from $\alpha$ by
deleting ``$,a_i,\vec{s}_i$'' whenever $a_i\notin\acts(A_k)$, and replacing the remaining pairs $\vec{s}_i$
by $\vec{s}_i[k]$.

Moreover, if $\sigma$ is a sequence of external actions of $A_1\|A_2$,
then $\sigma {\upharpoonright} A_k$ is what is left of $\sigma$
after removing all actions outside $\acts(A_k)$.
\end{definitionR}
Note that if $\sigma=\trace(\alpha)$, for $\alpha$ an execution of $A_1\|A_2$,
then $\sigma{\upharpoonright}A_k = \trace(\alpha[k])$.
Moreover, if $\alpha$ is an execution of $T\|A$, were $T$ is a test and $A$ a tested automaton,
then all rewards of the actions in $\alpha$ are inherited from the ones in $\alpha[1]$, so that
\begin{equation}\label{reward}
\reward(\alpha) = \reward(\alpha[1]).
\end{equation}

\begin{theoremR}{reward}
  $A  \mathop\sqsubseteq_{\rm reward}^{\it F} B$
  iff
  $A  \mathop\sqsubseteq_{\rm must}^{\it F} B$
  iff
  $A \mathop{\sqsubseteq_F} B$.
\end{theoremR}
 
\begin{proof}
That $A  \mathop\sqsubseteq_{\rm reward}^{\it F} B$ implies $A  \mathop\sqsubseteq_{\rm must}^{\it F} B$
has been shown in \cite[Thm.~7]{vG19b} and is also justified in \sect{testing}.

That $A  \mathop\sqsubseteq_{\rm must}^{\it F} B$ implies $A \mathop{\sqsubseteq_F} B$ has been
demonstrated by \thm{fair must testing}.

Suppose $A \mathbin\sqsubseteq_F B$, i.e., $\ina(A)\mathbin=\ina(B) \wedge \out(A)\mathbin=\out(B)$
and $\fairtraces(B) \mathbin\subseteq \fairtraces(A)$, and let $T$ be any test that is strongly
compatible with $A$ and $B$.\linebreak
Let $\beta$ be a fair execution of $T\|B$.
By \cite[Prop.~4]{LT89}, $\beta[1]$ is a fair execution of $T$, and $\beta[2]$ is a fair execution of $B$.
Since $A \sqsubseteq_F B$, automaton $A$ has a fair execution $\gamma$ with $\trace(\gamma)=\trace(\beta[2])$.
Let $\sigma:=\trace(\beta)$.
Then $\sigma$ is a sequence of external actions of $T\|A$ such that
$\sigma{\upharpoonright}T = \trace(\beta[1])$ and
$\sigma{\upharpoonright}A = \sigma{\upharpoonright}B = \trace(\beta[2]) = \trace(\gamma)$.
By \cite[Prop.~5]{LT89}, there exists a fair execution $\alpha$ of $T\|A$ such that
$\trace(\alpha)=\sigma$, $\alpha[1]=\beta[1]$ and $\alpha[2]=\gamma$.
By (\ref{reward}) one has $\reward(\alpha) = \reward(\alpha[1])=\reward(\beta[1])=\reward(\beta)$.
Thus $A  \mathop\sqsubseteq_{\rm reward}^{\it F} B$.
\qed
\end{proof}

\section{Conclusion}

When adapting the concept of a complete execution, which plays a central r\^ole in the definition
of must testing, to the weakly fair executions of I/O automata, must testing turns out to
characterise exactly the fair preorder on I/O automata. Moreover, reward testing, which under the
default notion of a complete execution is much more discriminating than must testing, in this
setting has the same distinguishing power. Interesting venues for future investigation include
extending these connections to timed and probabilistic settings.

\bibliography{references}
\end{document}